\documentclass[12pt]{article}
\usepackage{epsfig,amsfonts,amssymb}
\usepackage{hyperref}

\usepackage{cite}
\usepackage{cite}

\usepackage{comment}

\def\ZZZ{{\hbox{ Z\kern-1.6mm Z}}}
\def\RRR{{\hbox{ R\kern-2.4mm R}}}
\def\CCC{{\hbox{ C\kern-2.0mm C}}}
\def\zzz{{\hbox{z\kern-1mm z}}}

\def\ZZZ{{\mathbb Z} }
\def\RRR{{\mathbb R} }
\def\CCC{{\mathbb C} }

\newcommand{\qeq}{{\hbox{=\kern-2.3mm ? \kern.5mm }}}
\renewcommand{\qeq}{=}

\newcommand{\AAA}{{\cal A}}

\newcommand{\wt}{\widetilde}
\newcommand{\wh}{\widehat}

\newcommand{\NN}{{\cal N}}

\newcommand{\be}{\begin{equation}}
\newcommand{\ee}{\end{equation}}
\newcommand{\ben}{\begin{eqnarray}\displaystyle}
\newcommand{\een}{\end{eqnarray}}

\newcommand{\refb}[1]{(\ref{#1})}

\def\one{{\hbox{ 1\kern-.8mm l}}}
\def\zero{{\hbox{ 0\kern-1.5mm 0}}}

\newcommand{\bea}[1]{\begin{eqnarray}\label{#1} }
\newcommand{\eea}{\end{eqnarray}}

\newcommand{\eqref}{\refb}

\usepackage{bm}
\usepackage[table]{xcolor}

\def\NP{k}

\begin{document}

\baselineskip 24pt

\begin{center}

{\Large \bf Muti-instanton Amplitudes in Type IIB String Theory}


\end{center}

\vskip .6cm
\medskip

\vspace*{4.0ex}

\baselineskip=18pt

\centerline{\large \rm Ashoke Sen}

\vspace*{4.0ex}

\centerline{\large \it Harish-Chandra Research Institute, HBNI}
\centerline{\large \it  Chhatnag Road, Jhusi,
Allahabad 211019, India}


\vspace*{1.0ex}
\centerline{\small E-mail:  sen@hri.res.in}

\vspace*{5.0ex}

\centerline{\bf Abstract} \bigskip

We compute the normalization of the multiple D-instanton amplitudes in type IIB string theory and
show that the result agrees with the prediction of S-duality due to Green and Gutperle.

\vfill \eject

D-instanton contribution to the four graviton amplitude in type IIB string theory was predicted by
Green and Gutperle by requiring the amplitude to be 
S-duality invariant\cite{9701093,9704145,9808061}.
Direct computation of these amplitudes suffers from certain ambiguities related to integration
over zero modes. Recently these ambiguities were resolved using string field 
theory\cite{2104.11109} and the
resulting leading term in the one instanton contribution to the amplitude was shown to agree
with the predictions of \cite{9701093,9704145}. Our goal in this paper will be to extend
the results to multi-instanton amplitudes. We shall not attempt to make the paper self-contained,
but assume familiarity with the analysis of \cite{2104.11109} and freely use the results of that 
paper.

We begin by computing the normalization constant that multiplies the $\NP$-instanton
amplitude. As in \cite{2104.11109}, this is formally given by the exponential of the annulus diagram
multiplied by $i$. Taking into account the effect of the $\NP\times \NP$ Chan-Paton factor,
the annulus amplitude should be given by $k^2$ times the result for a single instanton.
Indeed the first part of our analysis will proceed exactly along this line.
We normalize the open string fields as:
\be
|\psi_o\rangle = \sum_a T^a \, |\psi^a_o\rangle, \qquad Tr(T^a T^b) =\delta^{ab}\, ,
\ee
where $|\psi^a_o\rangle$ is normalized in the same way 
as the open string field on a single D-instanton.
$T^a$'s are $\NP\times \NP$ hermitian matrices describing generators of $U(\NP)$, normalized so
that,
\be\label{e2}
Tr(T^aT^b)=\delta_{ab}\, ,
\ee
with $T^0=I_\NP/\sqrt \NP$ representing the $U(1)$ generator.
Proceeding as in \cite{2104.11109}, we get the
analog of eq.(4.32) of \cite{2104.11109}:
\be\label{enewernn}
 \NN_\NP=i\, \ (2\pi)^{-5\, \NP^2}\, (2\sqrt \pi)^{\NP^2}\, 
 \int \prod_{a=0}^{\NP^2-1} \left\{\prod_\mu d\xi_\mu^a \right\}  \left\{\prod_\alpha d\chi_\alpha^a\right\} 
e^S \Bigg/ \int\prod_{b=0}^{k^2-1} D\theta^b\, .
\ee
We have dropped the multiplier factor $\zeta$ that appeared in \cite{2104.11109} since there it was
shown to be unity.

The next step is to find the relation between the modes $\xi_\mu^0$ and the center of mass coordinate
$\wt\xi_\mu$ of the D-instanton system. For this we compare the effect of inserting a $\xi_\mu^0$ into
a disk amplitude of closed and open strings, with the closed strings carrying total momentum 
$p$, to the expected coupling $ip.\wt\xi$ of the center of mass coordinate. The only difference from
the computation in \cite{2104.11109} is that
the $\xi_\mu^0$ amplitude will get an extra factor of $1/\sqrt\NP$ due to the Chan-Paton factor 
$I_\NP/\sqrt\NP$. Trace over the Chan-Paton factors produces an extra factor of $k$, but this
affects the amplitudes with and without the $\xi_\mu^0$ insertions in the same way, 
and does not affect the ratio of the two amplitudes. Therefore
the analog of eq.(4.38) of \cite{2104.11109} takes the form:
\be
g_o\, \pi\, \sqrt 2\, \xi_\mu^0 / \sqrt \NP= \wt\xi_\mu\, ,
\ee
$g_o$ being the open string coupling on the D-instanton.

We also need to determine the relation between the gauge transformation parameters 
$\theta^a$ and the rigid $U(\NP)$ transformation parameters $\wt\theta^a$, defined so that
if $\wh\xi$ denotes a state of the open string with one end on the system of D-instantons 
under consideration and the other end on 
a spectator D-instanton, then the $U(\NP)$ transformation acts on
$\wh\xi$ as $e^{i\wt\theta_aT^a}\wh\xi$. 
Note that $\wh \xi$ in now a $\NP$ dimensional vector transforming in the fundamental 
representation of $U(\NP)$ since its one end can be attached to any of the $\NP$ D-instantons.
The analog of eq.(4.44) of \cite{2104.11109}, 
giving the infinitesimal string field theory gauge transformation of
$\wh\xi$, takes the form:
\be
\delta \wh\xi = {i\over 2} \, g_o \, \theta^a\, T^a \, \wh \xi\, .
\ee
Comparing this with the infinitesimal rigid $U(\NP)$ gauge transformation 
$\delta \wh\xi=i\wt\theta^aT^a\wh\xi$, we get,
 \be\label{ethetarange}
\theta^a = 2\, \wt\theta^a/g_o\, .
\ee
Since $\wt\theta^0$ accompanies the generator $T^0=I_\NP/\sqrt\NP$, it has period $2\pi\sqrt \NP$.
However since for $\wt\theta^0=2\pi/\sqrt \NP$, the $U(1)$ transformation coincides with the
$SU(\NP)$ transformation diag($e^{2\pi i/\NP}, \cdots, e^{2\pi i/\NP})$, once we allow 
$\wt\theta^0$ to span the full range $(0, 2\pi\sqrt\NP)$, the integration over $\wt\theta^a$'s for
$a\ge 1$
need to be restricted so that they span the group $SU(\NP)/\ZZZ_{\NP}$. This gives, using
\refb{ethetarange},
\be\label{e7}
\int\prod_b D\theta^b=2^{\NP^2}\, (g_o)^{-\NP^2}\, (2\pi \sqrt \NP)\, V_{SU(\NP)/\ZZZ_{\NP}}, \quad
V_{SU(\NP)/\ZZZ_\NP}\equiv \int_{SU(\NP)/\ZZZ_{\NP}}\prod_{a=1}^{\NP^2-1}D\wt\theta^a\, ,
\ee
where $\prod_a D\wt\theta^a$ is the Haar measure, normalized so that near the identity element the
integration measure is $\prod_a d\wt\theta^a$, and $\wt\theta^a$ are defined so that the $SU(\NP)$
matrix is given by $\exp[i\sum_{a=1}^{\NP^2-1} \wt\theta^a T^a]$.

As in \cite{2104.11109}, we further express $\chi^0_\alpha$ as
\be
\chi^0_\alpha = \wt\chi_\alpha/g_o\, ,
\ee 
so that the vertex operators of the modes $\wt\chi_\alpha$ do not carry any extra factor of $g_o$.
This gives
\ben\label{enewonenn}
 \NN_\NP&=& i\, \ (2\pi)^{-5\, \NP^2}\, 
 (2\sqrt \pi)^{\NP^2}\,\left({\sqrt \NP\over g_o\, \pi\sqrt  2}\right)^{10}\, 
  2^{-\NP^2}\, (g_o)^{\NP^2}\, {1\over 2\pi\sqrt \NP}\, {1\over V_{SU(\NP)/\ZZZ_{\NP}}}\, g_o^{16}
  \int\prod_{\mu=0}^9 d\wh\xi_\mu \prod_{\alpha=1}^{16} 
  d\wt\chi_\alpha \nonumber \\
&& \int \prod_{a=1}^{\NP^2-1} \left\{\prod_\mu d\xi_\mu^a \right\}  \left\{\prod_\alpha d\chi_\alpha^a\right\} 
e^S\, .
\een
For $\NP=1$ this reduces to the normalization constant given in eq.(4.49) of 
\cite{2104.11109}.

Let $\AAA_\NP$ denote the product of four disk amplitudes, each with one graviton and four 
$\wt\chi_\alpha$ insertions. The result takes the same form as in the case of one instanton 
amplitude, except that each $\wt\chi_\alpha$ is accompanied by a factor of $1/\sqrt\NP$ from
$T^0$, and each disk amplitude gives a factor of $\NP$ from trace over the Chan-Paton factors. 
This gives
\be
\AAA_\NP = \NP^{-8} \NP^4 \, \AAA_1= \NP^{-4}\, \AAA_1\, .
\ee
This gives the ratio of the coefficient of the $\NP$ instanton contribution to that of
the 1 instanton contribution to be:
\be \label{efirstint}
{\NN_k\, \AAA_k \over \NN_1\, \AAA_1} =(2\pi)^{-5\, (\NP^2-1)}(g_o\sqrt \pi)^{\NP^2-1}\, 
\sqrt \NP\, {1\over V_{SU(\NP)/\ZZZ_\NP}}\, 
\, \int \prod_{a=1}^{\NP^2-1} \left\{\prod_{\mu=0}^9 d\xi_\mu^a \right\}  
\left\{\prod_{\alpha=1}^{16} d\chi_\alpha^a\right\} 
e^S\, .
\ee

Next we shall determine the action $S$. The action vanishes up to quadratic order, but in order
to carry out the integration over the modes $\xi_\mu^a$ and $\chi_\alpha^a$ for $1\le a\le (\NP^2-1)$,
we need to keep higher order terms.\footnote{One did not need to do this in the analysis of
multi-instanton amplitudes in \cite{1912.07170}, since there was only one set of $\wt\xi_\mu^a$'s
associated with the Euclidean time direction, and  $\xi_0^a T^a$ was gauge equivalent to
a diagonal matrix.}
It is sufficient to consider the effective action obtained by
dimensional reduction of ten dimensional $N=1$ supersymmetric Yang-Mills theory to zero space-time
dimensions. Recalling that we have normalized the modes so that the (would be) kinetic terms 
are canonically normalized, the action takes the form:
\ben \label{eymaction}
S &=& {g_o^2\over 8} Tr\left( [A_\mu, A_\nu] [A^\mu,A^\nu]\right) 
+ {1\over 2\sqrt 2}\, g_o \gamma^\mu_{\alpha\beta} 
Tr\left(\Phi_\alpha 
[A_\mu, \Phi_\beta]\right), \nonumber \\
&&  A_\mu \equiv \sum_{a=1}^{\NP^2-1} 
\xi_\mu^a T^a, \quad \Phi_\alpha \equiv \sum_{a=1}^{\NP^2-1} \chi^a_\alpha T^a\, ,
\een
where we have taken into account the relation $g_o=\sqrt 2 \, g_{YM}$ between the open string
coupling constant $g_o$ and the Yang-Mills theory coupling constant 
$g_{YM}$\cite{polchinski}. One can
also check explicitly that the $\chi_\alpha^a$-$\chi_\beta^b$-$\xi_\mu^c$ amplitude computed
from \refb{eymaction} agrees with that computed from string theory in the convention of
\cite{2104.11109} and that the quartic coupling between the $\xi_\mu^a$'s agrees with the result of
\cite{1801.07607}.

The integral appearing in \refb{efirstint} is the partition function of the IKKT matrix 
model\cite{9612115}
and has been analyzed in \cite{9711073,9803117,9803265}.  
In particular, for general $\NP$, the result of this integral was conjectured in \cite{9803117}
and computed in \cite{9803265}.
Possible connection of this integral to the results of
\cite{9701093,9704145}  was also anticipated in
\cite{9711107,9803117,9903050}. Our main goal here will be to check that the integral 
\refb{efirstint} exactly reproduces the prediction of \cite{9701093,9704145} including the
normalization. To this end, 
we first define,
\ben \label{e13}
&&  x_\mu^a= g_o^{1/2} \, \xi_\mu^a\, , \quad y_\alpha^a = 
 g_o^{1/4} \, \chi_\alpha^a, \nonumber \\ &&
 X_\mu = x_\mu^a \, T^a =  g_o^{1/2} \, A_\mu, \quad Y_\alpha = y^a_\alpha\, T^a =
 g_o^{1/4} \Phi_\alpha
\, .
\een
This gives
\be \label{e14}
S = {1\over 8} Tr( [X_\mu, X_\nu] [X^\mu,X^\nu]) + {1\over 2\sqrt 2}\, \gamma^\mu_{\alpha\beta} 
Tr(Y_\alpha 
[X_\mu, Y_\beta])\, ,
\ee
and
\be\label{eratio}
{\NN_k\, \AAA_k \over \NN_1\, \AAA_1} =(2\pi)^{-5\, (\NP^2-1)}(\sqrt \pi)^{\NP^2-1}
\, \NP^{1/2}\, 
{1\over V_{SU(\NP)/\ZZZ_\NP}}\, 
\, \int \prod_{a=1}^{\NP^2-1} \left\{\prod_{\mu=0}^9 dx_\mu^a \right\}  
\left\{\prod_{\alpha=1}^{16} dy_\alpha^a\right\} 
e^S\, .
\ee

We shall now give the result of the integral appearing in \refb{eratio}
following the notation of \cite{9803117}.  
The action of \cite{9803117} took the form:
\be 
S =  {1\over 2} Tr\left( [\overline X_\mu, \overline X_\nu] [\overline X^\mu,\overline X^\nu]\right) 
+ \gamma^\mu_{\alpha\beta} 
Tr\left(\overline Y_\alpha 
[\overline X_\mu, \overline Y_\beta]\right)\, , 
\ee
with 
\be \label{e17}
\overline X_\mu = x_\mu^a \, \overline T^a \, , \quad \overline Y_\alpha = y^a_\alpha\, \overline T^a\, ,
\ee
with $\overline T^a$ normalized as
\be\label{e18}
Tr (\overline T^a \overline T^b) = {1\over 2} \delta_{ab}\, .
\ee
Comparing \refb{e18}, \refb{e17} with \refb{e2}, \refb{e13} we see that we have,
\be
\overline T^a ={1\over \sqrt 2} T^a, \qquad \overline X_\mu = {1\over \sqrt 2} X_\mu, \qquad
\overline Y_\alpha = {1\over \sqrt 2} Y_\alpha\, .
\ee
This gives
\be 
S =  {1\over 8} Tr( [X_\mu, X_\nu] [X^\mu,X^\nu]) + {1\over 2\sqrt 2}\, \gamma^\mu_{\alpha\beta} 
Tr(Y_\alpha 
[X_\mu, Y_\beta])\, ,
\ee
in agreement with \refb{e14}. The result of \cite{9803117} can now be stated as
\be \label{e21}
\int \prod_{a=1}^{\NP^2-1} \left\{\prod_{\mu=0}^9 {dx_\mu^a\over \sqrt{2\pi}} \right\}  
\left\{\prod_{\alpha=1}^{16} dy_\alpha^a\right\} 
e^S={2^{\NP(\NP+1)/2} \pi^{(\NP-1)/2} \over 2\sqrt{\NP} \prod_{i=1}^{\NP-1} i!}\, 
\sum_{d|\NP} {1\over d^2}\, .
\ee

We now turn to $V_{SU(\NP)/\ZZZ_\NP}$. The volume of $SU(\NP)$ was computed in
\cite{marinov} to be
\be
\wh V_{SU(\NP)} = {2^{(\NP-1)/2} \pi^{(\NP-1)(\NP+2)/2} \sqrt{\NP}\over \prod_{i=1}^{\NP-1} i!}
\, .
\ee
Ref.\cite{marinov} used algebra generators $\wh T^a$ normalized as
\be
Tr(\wh T^a \wh T^b) = 2\, \delta_{ab}\, ,
\ee
and labelling the group element as $\exp(i\wh \theta^a \wh T^a)$, defined the integration measure so
that near the origin the measure is $\prod_a d\wh \theta^a$. Comparing this with \refb{e2} and the
measure described below \refb{e7}, we see that we have $T^a =\wh T^a/\sqrt 2$, $\wt\theta^a =
\sqrt 2\, \wh\theta^a$ and $V_{SU(\NP)} = 2^{(\NP^2-1)/2}\wh V_{SU(\NP)}$. This gives,
\be\label{egrvol}
V_{SU(\NP)/\ZZZ_\NP} = V_{SU(\NP)}/\NP = 2^{(\NP^2-1)/2}\wh V_{SU(\NP)} /\NP
=  2^{(\NP^2-1)/2}
 {2^{(\NP-1)/2} \pi^{(\NP-1)(\NP+2)/2} \over \sqrt{\NP} \prod_{i=1}^{\NP-1} i!}\, .
\ee
Substituting \refb{e21}, \refb{egrvol} into \refb{eratio} we get,
\be
{\NN_k\, \AAA_k \over \NN_1\, \AAA_1} = \NP^{1/2} \sum_{d|\NP} {1\over d^2}\, .
\ee
This is in perfect agreement with the result of 
\cite{9701093,9704145}.

If we denote by $f(\tau,\bar\tau)$ the coefficient of the $R^4$ term in type IIB string theory action,
with $\tau=\tau_1+i\tau_2$ denoting the axion-dilaton modulus, and expand $f$ as $\sum_{k\in \ZZZ} f_k(\tau_2) e^{2\pi i k
\tau_1}$, then our analysis determines the coefficient of the leading term
in the large $\tau_2$ expansion of $f_k(\tau_2)$ for each $k$. On the other hand the requirement
of space-time supersymmetry gives a homogeneous 
linear second order  partial differential equation for 
$f(\tau,\bar\tau)$\cite{9808061,1502.03810}, 
which translates to a homogeneous  linear second order ordinary differential equation 
for each $f_k$.
Of the two solutions, one is unphysical since it has terms that grow exponentially in the large
$\tau_2$ limit. The other solution is determined uniquely once we determine the leading term in 
its large $\tau_2$ expansion. Therefore our result, together with supersymmetry, determines the
function $f(\tau,\bar\tau)$ completely without the help of S-duality.

\bigskip

\noindent {\bf Acknowledgement:} I wish to thank Rajesh Gopakumar,
Michael Green, Nobuyuki Ishibashi, 
Nikita Nekrasov and 
D Surya Ramana
for useful discussions. 
This work was
supported in part by the  Infosys chair professorship and the
J. C. Bose fellowship of 
the Department of Science and Technology, India.


\begin{thebibliography}{99}

\bibitem{9701093} 
  M.~B.~Green and M.~Gutperle,
  ``Effects of D instantons,''
  Nucl.\ Phys.\ B {\bf 498}, 195 (1997)
  doi:10.1016/S0550-3213(97)00269-1
  [hep-th/9701093].

\bibitem{9704145}
M.~B.~Green and P.~Vanhove,
``D instantons, strings and M theory,''
Phys. Lett. B \textbf{408}, 122-134 (1997)
doi:10.1016/S0370-2693(97)00785-5
[arXiv:hep-th/9704145 [hep-th]].

\bibitem{9808061}
M.~B.~Green and S.~Sethi,
``Supersymmetry constraints on type IIB supergravity,''
Phys. Rev. D \textbf{59}, 046006 (1999)
doi:10.1103/PhysRevD.59.046006
[arXiv:hep-th/9808061 [hep-th]].


\bibitem{2104.11109}
A.~Sen,
``Normalization of Type IIB D-instanton Amplitudes,''
[arXiv:2104.11109 [hep-th]].

\bibitem{1912.07170} 
  B.~Balthazar, V.~A.~Rodriguez and X.~Yin,
  ``Multi-Instanton Calculus in $c = 1$ String Theory,''
  arXiv:1912.07170 [hep-th].


\bibitem{polchinski}
J.~Polchinski,
``String theory. Vol. 1 and 2,'' Cambridge University Press,
doi:10.1017/CBO9780511618123

\bibitem{1801.07607}
C.~Maccaferri and A.~Merlano,
``Localization of effective actions in open superstring field theory,''
JHEP \textbf{03}, 112 (2018)
doi:10.1007/JHEP03(2018)112
[arXiv:1801.07607 [hep-th]].

\bibitem{9612115}
N.~Ishibashi, H.~Kawai, Y.~Kitazawa and A.~Tsuchiya,
``A Large N reduced model as superstring,''
Nucl. Phys. B \textbf{498}, 467-491 (1997)
doi:10.1016/S0550-3213(97)00290-3
[arXiv:hep-th/9612115 [hep-th]].

\bibitem{9711073}
T.~Suyama and A.~Tsuchiya,
``Exact results in N(c) = 2 IIB matrix model,''
Prog. Theor. Phys. \textbf{99}, 321-325 (1998)
doi:10.1143/PTP.99.321
[arXiv:hep-th/9711073 [hep-th]].

\bibitem{9803117}
W.~Krauth, H.~Nicolai and M.~Staudacher,
``Monte Carlo approach to M theory,''
Phys. Lett. B \textbf{431}, 31-41 (1998)
doi:10.1016/S0370-2693(98)00557-7
[arXiv:hep-th/9803117 [hep-th]].

\bibitem{9803265}
G.~W.~Moore, N.~Nekrasov and S.~Shatashvili,
``D particle bound states and generalized instantons,''
Commun. Math. Phys. \textbf{209}, 77-95 (2000)
doi:10.1007/s002200050016
[arXiv:hep-th/9803265 [hep-th]].

\bibitem{9711107}
M.~B.~Green and M.~Gutperle,
``D Particle bound states and the D instanton measure,''
JHEP \textbf{01}, 005 (1998)
doi:10.1088/1126-6708/1998/01/005
[arXiv:hep-th/9711107 [hep-th]].

\bibitem{9903050}
P.~Vanhove,
``D instantons and matrix models,''
Class. Quant. Grav. \textbf{16}, 3147-3164 (1999)
doi:10.1088/0264-9381/16/10/308
[arXiv:hep-th/9903050 [hep-th]].

\bibitem{marinov}
M.~S.~Marinov, 
``Invariant volumes of compact groups,'' J.~Phys.~A: Math. Gen., {\bf 13}, 
3357-3366 (1980); 
 ``Correction to ?Invariant volumes of compact groups,''
 J.~Phys.~A: Math. Gen., {\bf 14}, 543-544 (1981).

\bibitem{1502.03810}
Y.~Wang and X.~Yin,
``Constraining Higher Derivative Supergravity with Scattering Amplitudes,''
Phys. Rev. D \textbf{92}, no.4, 041701 (2015)
doi:10.1103/PhysRevD.92.041701
[arXiv:1502.03810 [hep-th]].



\end{thebibliography}
\end{document}